\documentclass[runningheads]{llncs}
\usepackage[T1]{fontenc}
\usepackage{graphicx}
\begin{document}
\title{The 2nd Workshop on Agile Practice \& Research: A Summary and Call For Research}
\titlerunning{The 2nd Workshop on Agile Practice \& Research}
%
\author{Karen Eilers\inst{1}\orcidID{0000-0002-4992-3522} \and
Michael Neumann\inst{2}\orcidID{0000-0002-4220-9641} \and
Eva-Maria Schön\inst{3}\orcidID{0000-0002-0410-9308} \and
Mali Senapathi\inst{4}\orcidID{0000-0003-3083-8069} \and
Maria Rauschenberger\inst{3}\orcidID{0000-0001-5722-576X} \and
Tiago Silva da Silva\inst{5}\orcidID{0000-0001-8459-7833} 
}
\authorrunning{K. Eilers et al.}
%
\institute{BSP Business and Law School - Campus Hamburg, Hamburg, Germany\\
\email{karen.eilers@bsp-campus-hamburg.de}
\and
University of Applied Sciences \& Arts Hannover, Hannover, Germany 
\email{michael.neumann@hs-hannover.de}\\
\and
University of Applied Sciences Emden/Leer,  Emden/Leer , Germany
\email{firstname.lastname@hs-emden-leer.de}\\
\and
Auckland University of Technology, Auckland, New Zealand\\
\email{mali.senapathi@aut.ac.nz}
\and
Federal University of S\~ao Paulo, S\~ao Paulo, Brazil\\
\email{silva.tiago@unifesp.br}}
\maketitle              
\begin{abstract}
Agile software development has been shaped by the interplay between academic research and industrial practice for over two decades, yet notable gaps persist between both domains. This paper focuses on three research–practice gaps: the theory gap, the time gap, and the transfer gap. To address these, the \textit{2nd Agile Practice \& Research Workshop} was held at the International Conference on Agile Software Development (XP) 2026 in S\~ao Paulo, Brazil, bringing researchers and practitioners together to identify root causes and develop joint solutions.
Building on two preceding sessions in which contributions of participants had been presented, participants engaged in a structured collaborative session, working in small groups on one of the three gaps and reflecting on possible causes and remedies. The organizers synthesized the results into four propositions for improving the research–practice intersection: (1) improving scientific communication, (2) aligning research more closely with emerging industrial needs, (3) creating stronger incentives for sustained collaboration, and (4) integrating educational approaches into research practice. From these, three calls for research were formulated: (a) broader adoption of open science practices for transparency, reproducibility, and cumulative evidence; (b) higher empirical quality standards through stronger theoretical grounding and rigorous design; and (c) more explicit, value-oriented contributions that clearly articulate their practical and scientific relevance. The paper offers both a summary of the workshop and a call to strengthen research–practice collaboration.

\end{abstract}
\section{Introduction}
Agile software development has attracted sustained attention in both research and practice for more than two decades~\cite{BahamHirschheim.2021}. During this time, a substantial body of knowledge has emerged on a wide range of topics, including agile methods and practices, scaling, requirements engineering, and human-centered software development. At the same time, agile practice has continuously evolved in response to changing organizational, technological, and societal conditions~\cite{EILERS.2022}. This ongoing evolution has increased both the need for and the difficulty of maintaining a strong connection between academic research and industrial practice.

Despite the maturity of the field, important gaps remain between agile research and practice. Practitioners often perceive research findings as difficult to access, insufficiently actionable, or only partially relevant to the complexity of real-world settings. At the same time, academic research frequently struggles to keep pace with the rapid development of industrial practices and emerging challenges~\cite{Barroca.2018}. To better understand and address this tension, it is useful to distinguish between different dimensions of the research-practice gaps in agile software development.

To contribute to this discussion, the \textit{2nd Agile Practice \& Research Workshop} was organized as part of the \textit{International Conference on Agile Software Development (XP) }2026. The workshop provided a platform for researchers and practitioners to share perspectives, discuss current challenges, and jointly reflect on how the relationship between agile research and practice can be strengthened. This paper summarizes the workshop and its main outcomes.

In this paper, we focus on three particularly relevant gaps that continue to shape the relationship between agile research and practice.

First, the \textit{theory gap} refers to the limited theoretical grounding of much agile research. Research in the agile community is still dominated by empirical case studies, which is common in interdisciplinary fields such as agile software development. However, practitioners often question the external validity of such studies and whether their findings can be generalized across the broader industry~\cite{Winters.2024}. In addition, literature~\cite{Yang.2025} points to theoretical limitations, such as a strong emphasis on framework practices rather than the underlying principles of the \textit{Agile Manifesto}~\cite{Beck.2001}. Although agile transformations and related phenomena have been widely studied (\textit{e.g.,} \cite {Dingsoeyr.2019,Carroll.2023}), there is still a need for stronger theoretical development that can explain, guide, and sustain agile practices across contexts~\cite{MagistrettiTrabucchi.2024,Schmid.2021}.

Second, the \textit{time gap} describes the challenge that industrial practice often evolves more rapidly than academic research. Over the past two decades, agile ways of working have continuously adapted to new organizational demands, technological developments, and shifting work environments~\cite{EILERS.2022}. Since 2020, this gap has become even more pronounced due to accelerated disruptive changes, including remote and hybrid work as well as the increasing integration of Artificial Intelligence (AI) into software development processes~\cite{Planoetscher.2025}. While agile teams in practice often show considerable resilience in responding to such developments, research frequently struggles to keep pace and to actively shape emerging trends. Addressing this imbalance requires closer alignment between the timescales of research and the realities of industry~\cite{Barroca.2018}.

Third, the \textit{transfer gap} captures the difficulty of making academic insights accessible, understandable, and actionable for practitioners. Although research generates valuable knowledge, its findings are often not communicated in ways that directly support decision-making and application in practice. Differences in language, priorities, and timescales between academia and industry further reinforce this challenge. Bridging the transfer gap, therefore, requires stronger collaboration between researchers and practitioners, as well as formats that translate research findings into applicable guidance for real-world agile settings~\cite{Woods.2025,Winters.2024}.
Together, these three gaps underscore the need for stronger academia–industry collaboration and form the conceptual foundation of the workshop summarized in this paper.

The paper is structured as follows. Section~\ref{sec:WorkshopOrg} outlines the workshop organization and thematic structure, followed by the results of the collaborative discussions. In Section~\ref{sec:Propositions}, we present our propositions to close the gaps before this summary closes with the calls for future research in Section~\ref{sec:Conclusion}.

\section{Workshop Organization}
\label{sec:WorkshopOrg}

Building on the results and findings of the first edition of the workshop in 2025~\cite{Neumann.2025a}, the second edition aimed to further strengthen the dialog between academia and industry. The workshop was designed as a half-day event within XP 2026 on 08.April~2026 in S\~ao Paulo, Brazil. Based on the three gaps, we invited contributions from both researchers and practitioners that addressed challenges and opportunities at the intersection of agile practice and research.
This year, we brought together 20 researchers and practitioners with diverse backgrounds, from Master's and PhD students to professors, and from consultants to software engineers and developers.

\subsection{Session 1: Bridging Time and Transfer Gaps Between Research and Practice}
\label{sec:Session1}

The first session focused on the challenges of aligning academic research with the pace and needs of industrial practice. Diebold~\cite{Diebold.2026} addressed this issue in his contribution \textit{Designing Transfer-Ready Research Artifacts for Ways of Working}, which discussed how research-based insights can be translated into practitioner-oriented artefacts that remain theoretically grounded while becoming easier to apply in real-world contexts. Walter et al. \cite{Walter.2026} further explored this challenge in \textit{Aligning Research and Product Development through Multi-Cycle Design Science Research: An Experience Report}, presenting a model for synchronizing academic research cycles with product development needs in industry. Contributions in this session therefore focused on solutions on how research can remain relevant in rapidly changing agile environments and how academic knowledge can be communicated in ways that are more accessible and actionable for practitioners. The discussion highlighted the need for more timely research, stronger exchange formats between academia and industry, and better mechanisms for translating findings into practical guidance.

\subsection{Session 2: Grounding Research for Practice: Closing the Theory Gap}
\label{sec:Session2}

The second session addressed the theory gap in agile software development research. Geger et al. \cite{Geger.2026} contributed with \textit{From Education to Evidence: A Collaborative Practice Research Platform for AI-Integrated Agile Development}, showing how project-based agile education can serve as a collaborative research platform for generating timely, practice-relevant evidence. Rook et al. \cite{Rook.2026} complemented this perspective with \textit{Closing Research--Practice Gaps Through Action Research: Empowering Teams to Strengthen Their Team Effectiveness}, which demonstrated how action research and theory-informed feedback loops can support teams in improving their ways of working while contributing to empirical theory development. Contributions in this session emphasized the importance of stronger theoretical grounding in order to move beyond isolated empirical observations and generate knowledge that is transferable across contexts. The discussion also raised the question of how theory development in agile research can be better connected to the realities of practice, so that conceptual advances remain both scientifically rigorous and practically relevant.

\subsection{Collaborative Session}
The collaborative session was designed to move beyond the presentation of individual contributions and to foster a structured discussion on how the identified gaps between agile research and practice can be addressed. To this end, the participants were divided into three small groups. To form the groups, the three tables at which the participants were initially seated were used as a starting point. Some participants were then asked to join a different group in order to ensure diversity with regard to participant characteristics (e.g., gender, age, and practitioner or researcher background) as well as to enable balanced group sizes. 

Each group was assigned one of the three previously identified gaps, namely the theory gap, the time gap, or the transfer gap. During the group work phases, the groups self-organized and documented the results of their discussions on sticky notes. The collaborative session comprised three small-group activities (15-minute timeboxes each) covering the following topics: 

\textit{(a) root causes of the gaps:} each group was asked to discuss and identify potential root causes underlying its assigned gap. Participants were encouraged not only to name possible causes but also to justify and critically reflect on them based on their own experiences from research and practice. 


\textit{(b) solution ideas to address the gaps:} the groups were invited to develop concrete solution ideas aimed at addressing the identified root causes and, consequently, contributing to closing the respective gap. 

\textit{(c) 15\% solutions:} each group identified measures that they could implement immediately on their own and without additional support or resources to realize the solutions developed in activity b). For this purpose, a method from Liberating Structures was applied. 

The time between the activities was used for plenary discussions of the individual group results, allowing participants to compare perspectives across groups and relate the proposed solutions to the broader challenges at the intersection of agile practice and research. For this purpose, the whiteboard walls in the workshop room were used to enable cross-group documentation of the results.


\subsection{Synthesis of the Workshop Results}
After the workshop, the insights generated during the collaborative session were synthesized by the attending workshop organizers (Author 1 and Author 2 of this summary). The synthesis was conducted immediately after the workshop in the same room. This enabled the documented results of the workshop participants to be used directly as input since they were still displayed on the whiteboard walls. In the first step, the identified root causes (activity a) and proposed solution ideas (activity b) were thematically categorized. In the second step, relationships between root causes and solution ideas were systematically analyzed in order to identify recurring patterns and overarching connections. Based on this synthesis, we derived a set of propositions aimed at improving the intersection between agile research and practice. In addition, we formulated specific calls for researchers intended to encourage the software engineering (SE) community to actively contribute to closing the identified gaps.

\section{Propositions for Optimizing the Intersection between Research \& Practice in Agile Software Development}
\label{sec:Propositions}

Based on the presented contributions, discussions conducted during the collaborative session and the subsequent synthesis of results, we derived four propositions to improve the intersection between research and practice in agile software development. These propositions reflect recurring themes across the identified root causes and solution ideas and point to areas in which more systematic change is needed.

\paragraph{Proposition 1: Scientific communication —}
A first proposition concerns the communication of scientific knowledge beyond academic audiences. 
A key barrier between research and practice is that practitioners often struggle to understand and apply research findings.
To address this issue, scientific communication skills should be more explicitly integrated into PhD study programs and other forms of research training. Researchers should be better prepared to present their work not only in academic settings but also in practitioner-oriented environments such as industry conferences, meetups, and community events. By doing so, researchers learn to identify the similarities between their own motivations and those of practitioners.  In addition, the SE community should experiment with alternative dissemination formats, including podcasts, vlogs, blogs, or similar channels. These formats could include, for example, new research results in practical language, instructional videos demonstrating how science and practice can collaborate, or success stories of previous collaborations in order to communicate evidence-based findings more broadly, more quickly, and more efficiently.

\paragraph{Proposition 2: Value the momentum —}
A second proposition emphasizes the importance of aligning research more closely with the topics and dynamics currently shaping industrial practice. To achieve stronger relevance, researchers need a greater awareness of those issues that practitioners perceive as urgent and valuable. In this context, the notion of value plays a central role. Research should more explicitly address practical concerns such as return on investment, business value, process-related pain points, and regulatory challenges. Furthermore, the discussions highlighted that strong networks in practice are essential for recognizing emerging topics early and understanding their relevance.  Building and maintaining such networks can therefore help researchers engage with practice at the right time and with greater impact.

\paragraph{Proposition 3: Emphasize incentives between research \& practice —}
The third proposition addresses the need for stronger incentives that encourage exchange and contribution between research and practice. One challenge lies in the absence of sufficiently visible and attractive mechanisms that motivate both communities to engage with one another. For practitioners, collaboration with academia may create opportunities to increase internal visibility, strengthen their influence within organizations, and contribute practical ideas to the broader scientific discourse. For researchers, closer interaction with practice can improve the relevance, visibility, and applicability of their work. Thus, more explicit incentive structures are needed that make the benefits of collaboration tangible for both sides and encourage sustained engagement across the research--practice divide.

\paragraph{Proposition 4: Integrating Educational Approaches into Research Practice —}
The fourth proposition focuses on the integration of educational concepts into research practice. The workshop discussions suggested that state-of-the-art educational methods, such as project-based learning, may provide valuable inspiration for the design and execution of research projects. Applying such approaches could help to structure research work more effectively, foster iterative learning, and sharpen the focus of project outcomes. In particular, this may support researchers in narrowing down research contributions and increasing the clarity and usefulness of outputs. Educational integration can therefore be understood as a way to strengthen both the process and the outcome orientation of research in contexts where relevance to practice is essential.

\section{Conclusion \& Call for Research}
\label{sec:Conclusion}
By synthesizing the results of this workshop, we also came up with specific calls for other researchers aiming to close the identified gaps:\\
\textbf{Call 1:} Establish a mandatory Open Science paradigm: To be able to increase transparency and thus understandability of research contributions, we call for a mandatory Open Science paradigm. Under the Open Science paradigm, we understand in accordance with Mendez et al. \cite{Mendez.2020} making research artifacts available to the public. The paradigm further expands to the openness of datasets themselves, data collection and analysis documentation, or open access, which can be ensured by publishing accepted papers as pre-prints. As the discussion on open science in SE goes back more than half a decade, we see the need to push this paradigm much further forward. Even if high quality SE venues like the \textit{International Conference on Software Engineering (ICSE)} or the \textit{International Symposium on Empirical Software Engineering and Measurement (ESEM)} have adopted mandatory Open Science policies, we notice that the majority of SE venues have not followed suit yet. We are aware of upcoming challenges and problems\textit{, e.g.,} regarding confidential datasets and open questions on how to deal with losing important contributions by enabling mandatory open science. But we see the need to move more into the direction of replicability and reproducibility of studies in the various different contexts to increase the chain of evidence in empirical SE research.\\
\textbf{Call 2:} Embrace the empirical research quality criteria towards high standards of rigorous, replicable and theory grounded research: Following-up on the initial call above, we want to  engage other researchers to focus even more on empirical research quality criteria and to push their own approaches more towards so-called gold-standard principles. We mean by that not only following standardized or established guidelines for preparing, conducting, and reporting research results. We further want to point to specific research quality criteria. Empirical SE research should be designed by a grounding theory. An analysis by Schmid~\cite{Schmid.2021} shows that empirical research focuses mostly on analyzing phenomena while not considering a theoretical grounding. By respecting quality criteria and grounding empirical SE research on a theory, we aim for replicable or reproducible research. In summary, this is what we call a gold standard for empirical SE research.\\ 
\textbf{Call 3:} Towards more explicit research contributions: Finally, we want to point out the very important facet of research contributions. The focus on contribution is typically grounded by providing arguments for expanding the existing body of knowledge with valuable additional findings. Although researchers are aware of this important aspect related to their research, it typically lacks one very specific point: What is the value of the results and/or findings? Furthermore, from a so called research gold standard claim, one has to consider that for research preparation and conduction in a systematic manner, the focus on contributions remains lacking. For example, Lago et al. \cite{Lago.2024} analyzed 91 ICSE 'best' papers between 2013 and 2024 and showed in their study that conclusion validity is the least considered threat in these studies. 

Interestingly, these findings show that we are not only challenged by the question of how we should explain or state such contributions but also how to critically reflect on these aspects as part of the limitations.

To summarize, from our point of view, research contributions are mainly divided into two categories: a) Artifacts grounded by research findings or results and b) Artifacts grounded in research methodologies. To be more precise, we provide two specific examples. As an example of an artifact grounded in research findings, we refer to the framework ``\textit{Genau Meine Arbeitsweise}''~\cite{Diebold.2026} discussed by the workshop participants. The value of this framework lies in enabling practitioners to reflect on their own ways of working and to adapt them in a more sustainable and deliberate manner. In this sense, the artifact translates prior empirical findings on work practices, reflection, and organizational change into an applicable instrument for practice. An example of an artifact grounded in research methodologies is the platform \textit{"Columinity.com"}~\cite{VerwijsRusso.2023}, presented and discussed in the workshop. During its use, the platform supports the systematic collection of scientific data while simultaneously guiding teams in applying Action Research principles in their own organizational context. Hence, the artifact creates value not only through its practical functionality but also by embedding a rigorous research methodology into real-world practice and thereby closing research-practice gaps. We provide additional material from the workshop in our git repository~\cite{Neumann.2026a}.

%
%
%
 \bibliographystyle{splncs04}
 \bibliography{references}

\end{document}